\documentclass[prb,aps,twocolumn,eqsecnum,superscriptaddress,showpacs,showkeys]{revtex4}
\usepackage{graphicx}
\usepackage{color}

\begin{document}

\title{
On the ferromagnetic character of (LaVO$_3$)$_m$/SrVO$_3$ superlattices} 
\author{Cosima Schuster}
\affiliation{Universit\"at Augsburg, Institut f\"ur Physik, 86135 Augsburg, Germany}
\author{Ulrike L\"uders}
\affiliation{Laboratoire CRISMAT, UMR CNRS-ENSICAEN 6508, 14050 Caen, France}
\author{Udo Schwingenschl\"ogl}
\affiliation{KAUST, PSE Division, Thuwal 23955-6900, Kingdom of Saudi Arabia}
\author{Raymond Fr\'esard}
\affiliation{Laboratoire CRISMAT, UMR CNRS-ENSICAEN 6508, 14050 Caen, France}

\begin{abstract}
The experimental observation that vanadate superlattices
(LaVO$_3$)$_m$/SrVO$_3$ show ferromagnetism up to room temperature [U.\
L\"uders {\it et al.}, Phys.\ Rev.\ B {\bf 80}, 241102R (2009)] is
investigated by means of density functional theory. First, the
influence of the density functional on the electronic and magnetic structure of 
bulk ${\rm LaVO_3}$ is discussed. Second, the band structure of a 
(LaVO$_3$)$_m$/SrVO$_3$ slab for $m=5$ and 6 is calculated. Very different
behaviors for odd and even values of $m$ are found: In the odd case lattice
relaxation results into a buckling of the interface VO$_2$ layers that leads to
spin-polarized interfaces. In the even case a decoupling of the interface
VO$_2$ layers from the LaO layers is obtained, confining the interface electrons
into a two-dimensional electron gas. The orbital reconstruction at the
interface due to the lattice relaxation is discussed. 
\end{abstract}


\maketitle

\section{Introduction}

The combination of low dimensionality and strong correlations proved fruitful
in the quest for new materials with functional-oriented properties. It lead to
the high-$T_c$ superconductivity,\cite{Malozemoff05,Bed86} 
transparent conducting oxides,\cite{Kawa97} 
high capacitance heterostructures,\cite{Li11} and large thermopower,\cite{Maignan09} 
to quote a few. Recently, the creation of high-mobility electron gases with
reduced dimensionality has also been pursued in artificially layered oxide
systems. They include (i) the interface between the two insulators 
${\rm Sr Ti O_3}$ and ${\rm La Al O_3}$,\cite{Ohtomo04,thiel06,Shalom10} (ii)
systems with ultra-thin doping layers as ${\rm Nb: Sr Ti O_3}$ in  
${\rm Sr Ti O_3}$,\cite{Kim11} and RO layers with
R = La, Pr, Nd  in  ${\rm Sr Ti O_3}$,\cite{Jang11} and (iii) the surface of
${\rm Sr Ti O_3}$.\cite{Santander} In some of these systems
functional properties were found, such as superconductivity\cite{rey07} and
ferromagnetism.\cite{Lib11}

Charge carriers with reduced dimensionality can be introduced in other complex
oxides, too. For instance, ultra-thin ${\rm Sr V O_3}$ layers can be
intercalated in ${\rm La V O_3}$,\cite{David11} resulting in ferromagnetism
up to room temperature\cite{Lueders} although none of the two oxides 
is magnetic at this temperature. This points out the relevance of the layered
geometry, since the solid 
solution La$_{1-x}$Sr$_x$VO$_3$ shows the typical phase diagram of a doped
Mott-insulator with a transition from an antiferromagnetic insulator to a
non-magnetic metal at about $x = 0.2$.\cite{Inaba,Miyasaka00} Contrary to the
solid solution, (LaVO$_3$)$_m$/SrVO$_3$ superlattices show ferromagnetic
behavior above room temperature for $m$ ranging from 3 to 6, with a higher
magnetization for even $m$ values than for odd values. Thus, the geometrically
confined doping in the superlattices gives rise to a magnetic phase which is
not observed in the randomly doped solid solution. 

From the theoretical point of view, the reduced dimension of electronic
states confined in a potential created by the interface of two insulators, or
by geometrically confined doping, leads to a renewed interest in 
two-dimensional systems. Studies of the three band Hubbard model predict a
ferromagnetic ground state in case of 2/3 and 3/4 band filling for a reduced
bandwidth due to the enhanced correlations of the electrons.\cite{Chan09}
Ferromagnetism was also theoretically predicted for a model specific to two
t$_{2g}$ orbitals,\cite{Fresard02} at SrTiO$_3$/LaVO$_3$
interfaces\cite{Jackeli08} and in SrTiO$_3$/SrVO$_3$
superlattices,\cite{pickett10} indicating a strong coupling of the electronic,
magnetic and orbital degrees of freedom to the geometry of the superlattice
including the number of SrVO$_3$ layers. Again, in both cases the magnetic
phase was largely explained by the confinement and a change in the orbital
occupations. 

However, in the case of superlattices or heterostructures, the reduced bandwidth
and therefore enhanced correlations are just one part of the story. As in
those systems two different materials are epitaxially grown on top of each
other, also the misfit of the lattice parameters of both materials plays a
role for the properties of the system mediated by the distortion of the
structure due to strain. Especially in oxides the distortion can have a strong
influence on the properties since, as mentioned above, the electronic, spin
and lattice degrees of freedom are strongly coupled. This effect was shown
theoretically in distorted LaVO$_3$ where, depending on the
distortion of the lattice, different antiferromagnetic structures become
energetically favorable.\cite{weng10} 

In order to analyze the ferromagnetic state observed in
(LaVO$_3$)$_m$/SrVO$_3$ superlattices, electronic band structure
calculations are helpful to confirm the confinement of the electronic states
and to elucidate the origin of the ferromagnetism. For this purpose, we will use
density functional theory and address the distortion of the
lattice. Furthermore, the superlattices show an asymmetry of the magnetic 
properties between odd and even numbers of LaVO$_3$ unit cells. Thus, one
aim of the calculations is to provide evidences for a possible difference in
confinement between the two geometries. 

\begin{figure*}[t!]
\begin{center}
\unitlength=0.18cm
\begin{picture}(44,72)
\put(-21, 35){\includegraphics*[width=6.7cm]{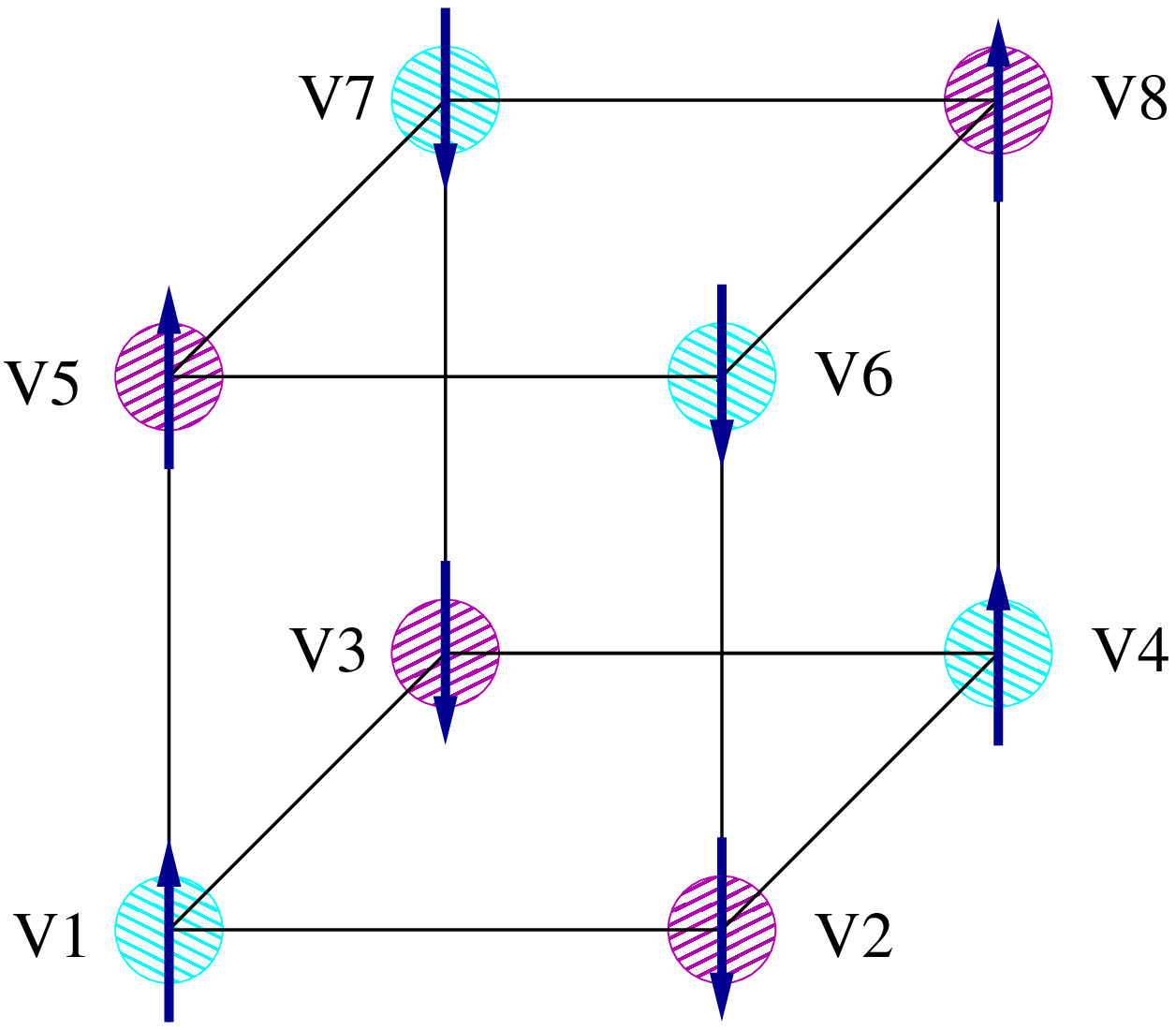}}
\put(29,35){\includegraphics*[width=6.7cm]{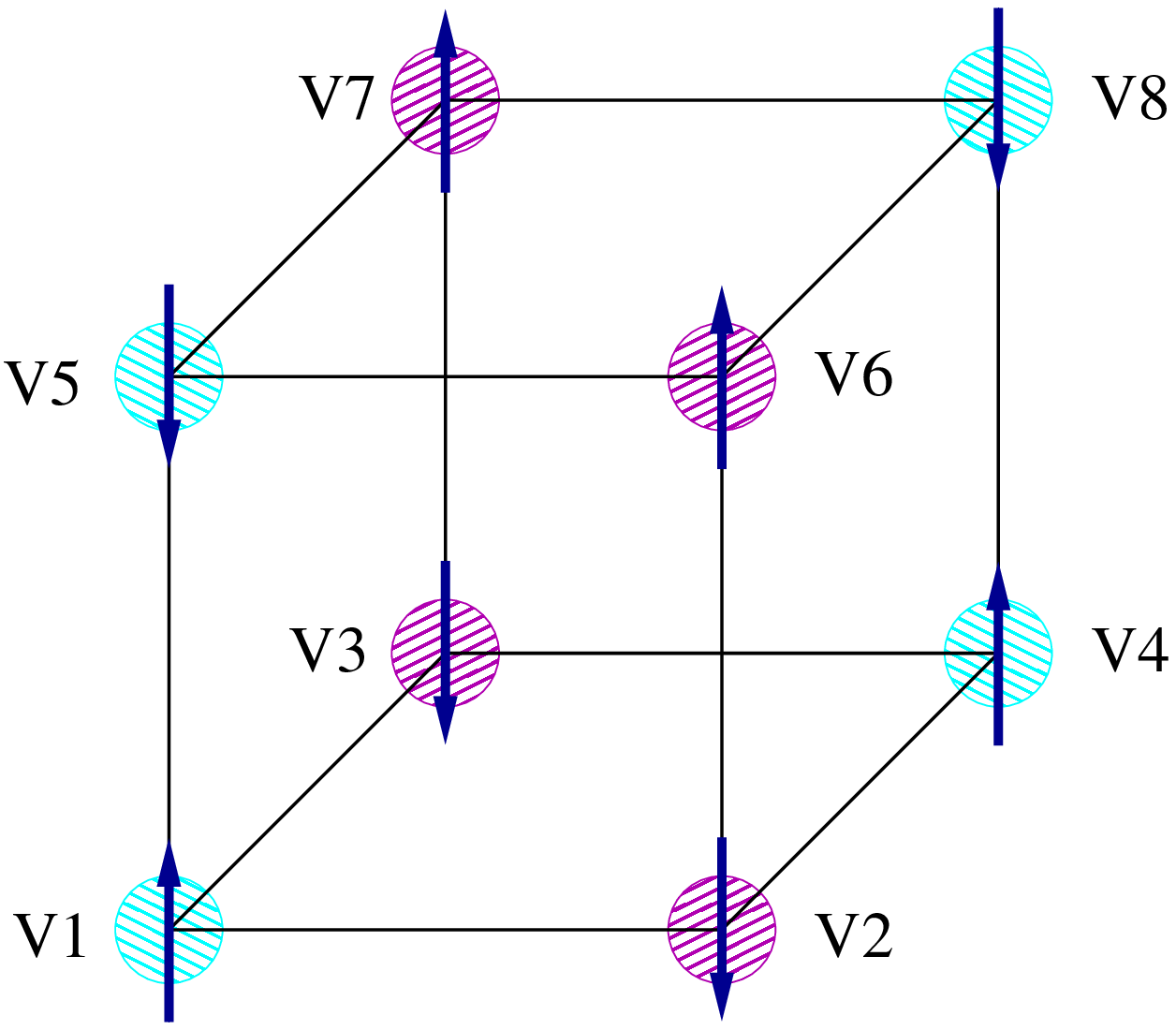}}
\put(-21, -2){\includegraphics*[width=6.7cm]{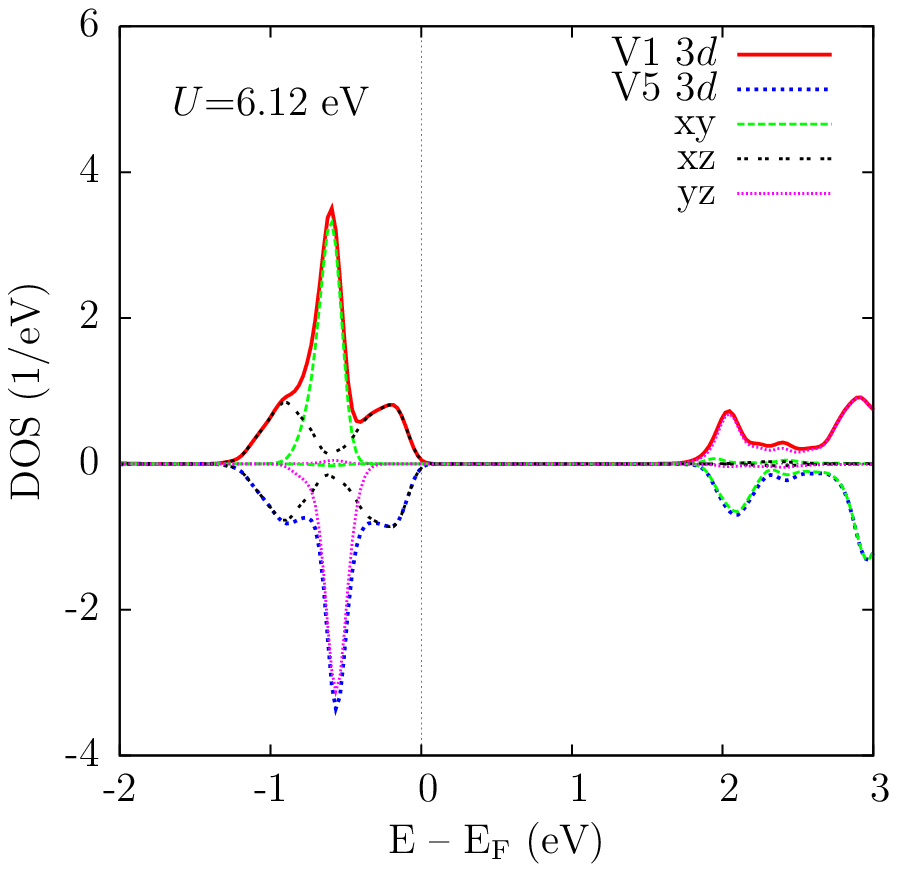}}
\put(29, -2){\includegraphics*[width=6.7cm]{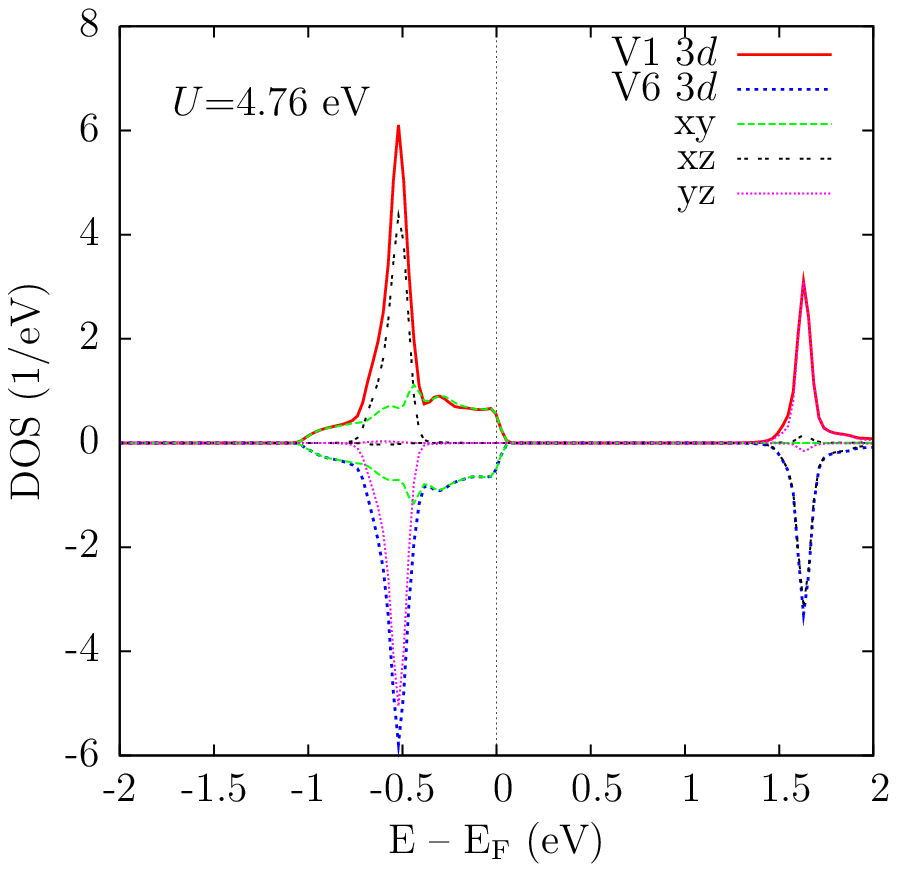}}
\end{picture}
\end{center}
\label{fig1}
\caption{(Color online) Top: Proposed spin and orbital order in the rare earth
vanadates. Bottom: Density of states projected on the V $3d$ orbitals.
Left: C-type spin ordered ${\rm LaVO_3}$. % $U$=6.12 eV; 
Right: G-type spin ordered ${\rm LaVO_3}$. % $U$=4.76 eV. 
The orbital occupancy of the V1 atom in the left lower corner (red line) is
compared to the orbital occupancy of the according V of the upper layer with
the same spin orientation (blue line). The latter DOS is multiplied by $-1$
and plotted along the negative y-axis. The $c/a$ ratio is 1.02.}
\end{figure*}

The paper is organized as follows. In Sec.~\ref{sec:setup} we discuss the
influence of the density functional on the electronic and magnetic structure of 
bulk ${\rm LaVO_3}$. The structure model of (LaVO$_3$)$_m$/SrVO$_3$
superlattices is addressed in Sec.~\ref{sec:sl}, followed by the electronic
structure, the charge carrier distribution and the orbital occupancy of the
$m=5$ and $m=6$ superlattices, as representative of even and odd
superlattices, respectively. Finally, our findings are summarized in
Sec.~\ref{sec:concl}.   

\section{Electronic and magnetic structure of bulk ${\rm LaVO_3}$}\label{sec:setup}
There are many systems for which there is a consensus about the appropriate
density functional. This is not the case for the vanadates 
and we therefore start our considerations addressing this point.
The rare earth vanadates with perovskite structure, including LaVO$_3$,
have the V ion in a $3+$ valence state, i.e., the two electrons in
the V $3d$ shell occupy two of the $t_{2g}$ orbitals with $S=1$. The
crystal field splitting of the $3d$ orbitals in threefold degenerate
$t_{2g}$ and twofold degenerate $e_g$ orbitals is a consequence of the
octahedral coordination of the V ions by O atoms. The rare earth
vanadates show simultaneous antiferro-type spin and orbital ordering.\cite{fujika} 
The spin ordering of LaVO$_3$ is C-type, i.e., the alignment is
antiferromagnetic within the $xy$-plane and ferromagnetic along the
$z$-direction.\cite{Zubkov} The accompanying orbital order is G-type, i.e., it
alternates in all directions. Other vanadates show G-type spin ordering
together with C-type orbital ordering. For an overview of the 
phase diagram see Ref.~\onlinecite{fujika}. It has been shown in
Ref.~\onlinecite{weng10} that strain due to the coherent growth on a substrate
can strongly affect the orbital ordering. For an increasing $c/a$ ratio the G-type order is
destabilized.

It is essential that the approximation to the density functional
treats the lattice, magnetic, and electronic properties on an equal
level. While the local density approximation (LDA) typically
underestimates the bond lengths and magnetic couplings, the generalized
gradient approximation (GGA) overestimates both. For bulk LaVO$_3$, the
GGA correctly predicts the C-type spin ordering, while the LDA tends
to delocalize the electronic states, which is critical in the case of
narrow $3d$ bands. To obtain an insulating state, corrections
beyond the LDA/GGA have to be considered, such as the LDA/GGA+$U$
approach in which a local interaction is added for the $3d$ orbitals.
\cite{Anisimov} This enhances the separation of the $3d$ energy levels
and opens a band gap. It is found that the magnetic ordering can
be changed by this treatment.\cite{schwin07}

A further step towards
a rigorous treatment can be achieved by hybrid functionals which
incorporate a non-local Fock exchange. Hybrid functionals such as
PBE0 and HSE yield an electronic structure comparable to self-consistent
GW calculations for much lower computational cost. Even the local EECE
approach has shown clear advantages over the LDA/GGA+$U$ in describing 
the electronic structure in nickelate stripe phases.\cite{nickelates}
However, it is known that the Fock term (without correction of the
correlation term) results in a substantial overestimation of the
magnetic coupling  and, therefore, in an unreliable
determination of the magnetic ordering, including phase transitions.
A GW treatment leads to accurate electronic structures and magnetic
phase transitions but is computationally too costly for superlattice
systems. The best compromise therefore is the LDA/GGA+$U$ approach, which we
apply in the following. Specifically, we employ the WIEN2k augmented plane
wave plus local orbitals code.\cite{wien2k} We find that the energies 
and electronic states obtained for the fully localized and around mean
field double counting corrections show the same behavior.

The interaction parameter $U$ is adjusted as follows. For
$U>5$ eV the C-type spin ordering is energetically favored over the
G-type ordering. A band gap of 1.5 eV is obtained for $U=6.12$ eV in
the case of C-type spin ordering, while the same gap is found in the case of
G-type spin ordering for $U=4.8$ eV, see Fig.~1. In addition, the C-type orbital
ordering under G-type spin ordering depends only weakly on $U$ and on the
lattice strain. On the other hand, the G-type orbital ordering under
C-type spin ordering is very sensitive to strain. Disorder or ferro-type
orbital ordering is found for $c/a>1$ (which is fulfilled in our
superlattice system). Ferromagnetism and A-type antiferromagnetism
result in higher energies than the C- and G-type spin orderings for all
approximations under study. Electronic densities of states for C-type
and G-type spin ordered LaVO$_3$ for $c/a=1.02$ are shown in Fig.~1. While the
G-type density of states (DOS) indicates very narrow bands, the bands
are much broader for C-type spin ordering despite the higher $U$ value.
For C-type spin ordering, the $d_{xz}$ orbital, forming a double peak,
is occupied, while the $d_{xy/yz}$ orbitals are partially occupied and
therefore subject to orbital ordering, in agreement with
Ref.~\onlinecite{weng10}. In contrast, for G-type spin 
ordering, the $d_{xy}$ orbital is occupied, with a much less pronounced
DOS structure, and the orbital ordering affects the $d_{xz/yz}$ orbitals.

\begin{figure}[b]
\includegraphics*[width=0.31\textwidth,angle=-90]{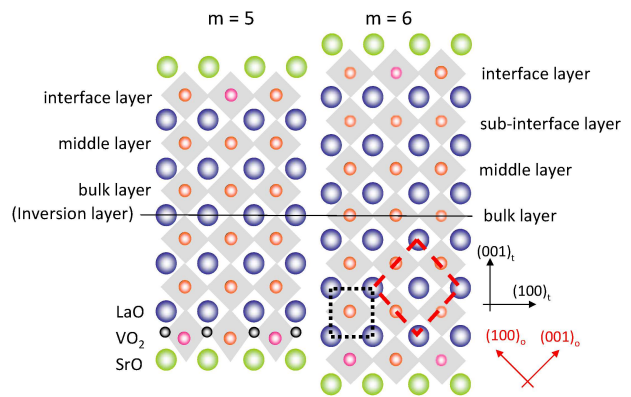}
\caption{(Color online) Supercell setup for the (LaVO$_3$)$_m$/SrVO$_3$
heterostructure for $m=5$ (left) and $m=6$ (right). La (Sr) atoms are
represented by blue (green) spheres and V$^{3+}$ (V$^{2+}$) ions are
represented by orange (pink) spheres. The grey squares indicate the oxygen
octahedra. The oxygen atoms themselves are omitted, except for the bottom
interface layer for $m=5$ where black spheres are shown to highlight the
buckling of the VO$_2$ layer. In the bottom of the $m=6$ structure the (101)
plane of the tetragonal (orthorhombic) cell is indicated in black (red).}
\label{fig2}
\end{figure}

Within GGA the ground state of bulk SrVO$_3$ is antiferromagnetic
with $\mu_V=0.2$ $\mu_B$, but the energy difference to the non-magnetic
solution is small, amounting to 0.2 meV per V atom.\cite{inoue} For the
GGA+$U$ method, the C-type antiferromagnet is favored by 30 meV per V
atom over the A-type antiferromagnet and by 140 meV per V atom over
the ferromagnet. A local hybrid approach with 25\% Fock mixing would favor
G-type spin ordering, since a C-type starting configuration changes to
G-type during the self-consistency calculation. The onsite interaction does
not open a band gap in bulk SrVO$_3$ but the compound stays metallic. Around
the Fermi energy, we find the $t_{2g}$ orbitals equally partially occupied,
while the $e_g$ orbitals are empty. Even though the spin-orbit coupling has
little impact on the ground state of LaVO$_3$ it must be included in the case
of SrVO$_3$.\cite{pickett10} Indeed it splits the $t_{2g}$ states and gives
rise to a band gap of 0.61 eV. In this case C-type spin ordering
(magnetization in the $x$-direction) comes along with filled $d_{xy}$ and
half-filled $d_{xz,yz}$ orbitals, yielding ferro-type orbital ordering.

\begin{figure*}
\begin{center}
\unitlength=0.18cm
\begin{picture}(44,36)
\put(-21, -3){\includegraphics*[width=6.7cm]{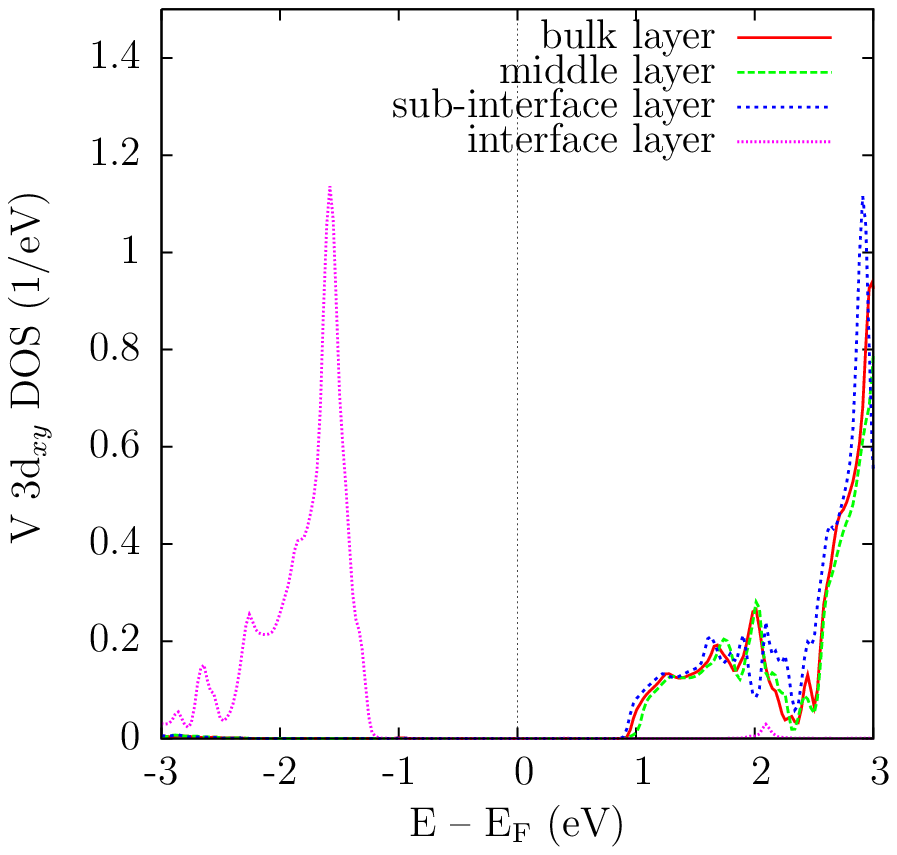}}
\put(29,-3){\includegraphics*[width=6.7cm]{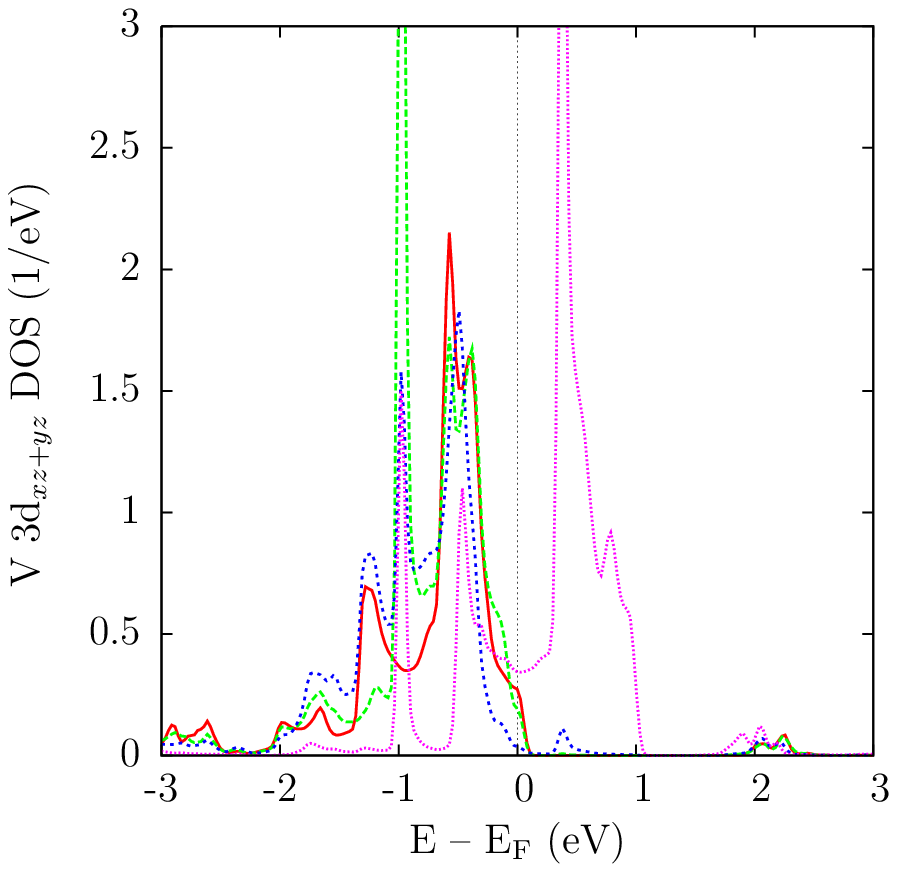}}
\end{picture}
\end{center}
\caption{(Color online) Density of states projected on the (left) V $3d_{xy}$
and (right) V $3d_{xz,yz}$ orbitals in the $m=6$ superlattice. The inversion layer
is denoted as bulk layer. The labeling of the layers is shown in Fig.~\ref{fig2}.} 
\label{fig6}
\end{figure*}

\section{(${\rm LaVO_3}$)$_m$/(${\rm SrVO_3}$) superlattices}\label{sec:sl} 

The structure setup for the calculations of the superlattice starts from the
tetragonal phase of LaVO$_3$, neglecting an orthorhombic lowering of the
symmetry due to rotations of the oxygen octahedra. In addition, ideal
atomically sharp interfaces are studied. We put one SrVO$_3$ layer between
$m=5$ or $m=6$ LaVO$_3$ layers, representative of the odd and even
superlattices, respectively. The lattice constants are $a=3.88$ \AA\  and
$c=(m+1)3.95$ \AA.  

As observed experimentally, the $c/a$ ratio in the superlattice deviates from
the bulk ratio of LaVO$_3$.\cite{David11} In bulk LaVO$_3$ it is smaller than 1,
$c/a=0.98$,\cite{Bordet93} but the strain exerted by the SrVO$_3$ layers
distorts the LaVO$_3$ sublayers in the superlattice. In case of the odd
superlattice, six VO$_2$ layers lie in between the SrO layers and the LaO
forms the inversion layer, while in the even superlattice with its seven
VO$_2$ layers, the central VO$_2$ layer is the inversion layer (see Figure~\ref{fig2}).  

In a first step, we neglect the antiferromagnetic order in the $ab$-plane and 
restrict ourselves to a $1\times1$ supercell in the $ab$-plane. 
In $c$-direction, ferromagnetic alignment is proposed, as expected for C-type
spin order. It turns out in a second step that when taking into account also
the magnetic ordering in the $ab$-plane by a $2\times2$ supercell the results
concerning the interface electronic structure do not depend on the ordering in
the $ab$-plane.

\subsection{Superlattice $m=6$}\label{sec:m6}

Let us start our investigations with the $m=6$ superlattice and first focus on
the orbital occupancy reconstruction at the interface.
Deep in the LaVO$_3$ region a V$^{3+}$ valency is realized, with the
$3d_{xz}$ and $3d_{yz}$ orbitals occupied and the V $3d_{xy}$ orbitals empty, as
revealed by Fig.~\ref{fig6}. On the contrary, for the interface layers (the
two VO$_2$ layers adjacent to the SrO layer) the V $3d_{xy}$ orbital is
occupied and the V $3d_{xz,yz}$ orbitals are partially occupied, see
Fig.~\ref{fig6}, together with a small admixture of the $e_g$ orbitals. This
results in a smaller valency of the V ions, which shows 
that they mostly retain the oxidation state of the respective bulk
material. This provides us with a first indication that the doping of LaVO$_3$
is geometrically confined. 

\begin{figure}[!b]
\includegraphics[width=0.45\textwidth]{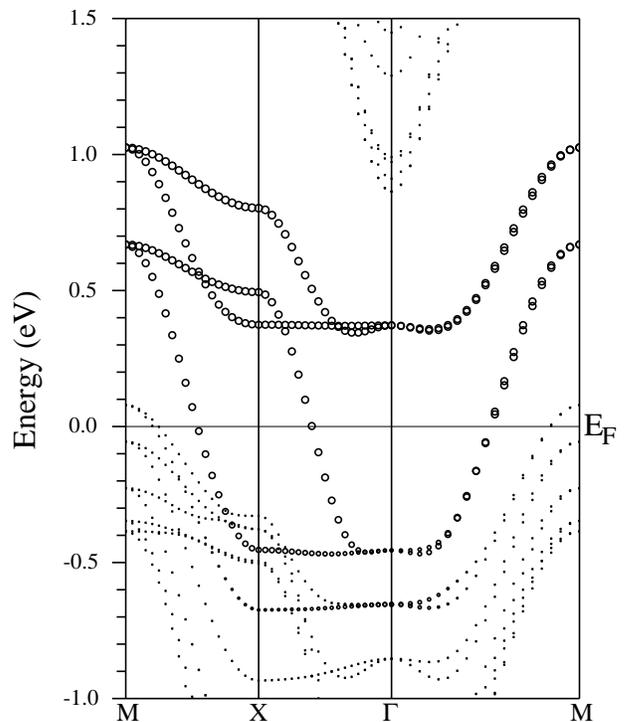} 
\caption{Band structure of the $m=6$ superlattice with non-magnetic
interfaces. The interface V $3d_{xz}$+$3d_{yz}$ orbitals are highlighted.} 
\label{fig7}
\end{figure}

This statement is further confirmed by the band structure shown in
Fig.~\ref{fig7}. It exhibits four interface-based bands crossing the Fermi
energy, two of them giving rise to fairly symmetrical Fermi surfaces (in the
$k_xk_y$-plane) centered around the $\Gamma$ point. These bands are
characterized by rather large Fermi velocities. In addition, small hole
pockets centered around the M-point are found. They are characterized by
clearly smaller Fermi velocities and are due to bulk states. Therefore, the
fast charge carriers are mostly confined to the interface layers, though in
states also containing a small admixture from the sub-interface layer. 
Additional features of the band structure point to an interesting
two-dimensional character of the bands crossing the Fermi energy. Principally
constituted by d$_{xz}$ (d$_{yz}$) orbitals which disperse along the
$\Gamma$-X ($\Gamma$-Y) direction, they unexpectedly show considerable
dispersion along X-M (Y-M), pointing to an admixture of other orbitals to the
d$_{xz}$ (d$_{yz}$) orbitals. Also, the dispersion around the $\Gamma$ point
is rather flat and not cosine-like as one would be expected for pure d$_{xz}$/d$_{yz}$ bands.

A partial occupation of the d$_{xz}$ and d$_{yz}$ orbitals establishes the
prerequisite of an orbitally ordered state. However, this is neither observed
in our calculations nor in the experiments, where it was shown that these
systems resemble rather good metals at low temperature.\cite{David11} An
explanation for this behavior may be found in the observed structure
relaxations, where the most important ones are summarized in table
\ref{tab:1}. In the interface layers the V (O) ions move towards (off) 
the SrO layers, resulting in a sizeable buckling of the interface layers
with an amplitude close to 0.11 \AA. The first LaO layers move quite
homogeneously towards the SrO layer, with a buckling amplitude smaller than
0.01 \AA, while the buckling amplitude of the sub-interface VO$_2$ layer is
close to 0.04 \AA.

\begin{table}[b!]
\caption{
Displacements (in \AA) off SrO layers of the V and O atoms in the interface and
middle/sub-interface layers for the heterostructures with $m=5$ and $m=6$. In
the first LaO layers for $m=5$ ($m=6$) the motion towards the SrO layer of the
La ions amounts to 0.118~\AA\ (0.093\ \AA), and that of the O ions to 0.116~\AA\
(0.010\ \AA).}
\vskip .1cm
\begin{ruledtabular}
\begin{tabular}{lcccc}
\multicolumn{1}{c}{}         &
\multicolumn{2}{c}{$m=5$}    &
\multicolumn{2}{c}{$m=6$}         \cr
\multicolumn{1}{c}{}         &
\multicolumn{1}{c}{V}           &
\multicolumn{1}{c}{O}           &
\multicolumn{1}{c}{V}           &
\multicolumn{1}{c}{O}         \cr
\colrule
Interface layer  & $-0.022$  & 0.096 & $-0.038$  & 0.076 \\
Middle/sub-interface layer & $-0.0046$  &  0.0029 & $-0.0027$ &0.0159\\
\end{tabular}
\end{ruledtabular}
\label{tab:1}
\end{table}
 
As a result, the bond lengths are strongly reduced between the V ions in the
interface VO$_2$ layers and the O ions in the SrO layer, becoming smaller than
$a/2$, i.e., the distance of the V ion to the O ions in the $xy$-plane of
the sample positioned at $a/2$ is larger than that to the O ion in the
$z$-direction. Consequently, the non-degenerate V energy level
$\varepsilon_{xy}$ lies below the $\varepsilon_{xz,yz}$ levels in the
interface layer, leading to an enhancement of the dispersion in those
layers. On the other hand, the enlarged bond length between the interface
layer and the adjacent LaO layer suppresses the orbital overlap in the
$z$-direction and favors therefore an electronic decoupling of the interface
VO$_2$ layer with respect to the other VO$_2$ layers, as shown in Figure~\ref{bond}.
The same tendency of the buckling can still be seen in the sub-interface layer, but with
a much smaller amplitude. The O-V bond length in this case is near to $c$/2 as for the
other VO$_2$ layers. Thus, the larger distance between the V ions and the
surrounding O atoms is to be found along the $z$-direction, inducing the bulk
orbital occupancy and, hence, an orbitally ordered insulating state like in the bulk. 

Thus, the calculated band structure accounts for a two-dimensional metallic
state, but it cannot account for the experimentally observed ferromagnetism. In a Stoner
picture this would be understood as the result of a rather small DOS at the Fermi
energy. Therefore, should this system be ferromagnetic then ferromagnetism
would likely be caused by further correlation effects, as taken into account
in calculations of multi-orbital Hubbard models.\cite{Fresard02,Chan09}
Moreover, we have verified that the above electronic structure is stable
against the employed $U$ value and lattice relaxation, as is the establishment
of an ultrathin two-dimensional electron gas. 

\subsection{Superlattice $m=5$}\label{sec:m5}

\begin{figure}[!t]
\includegraphics*[width=0.4\textwidth]{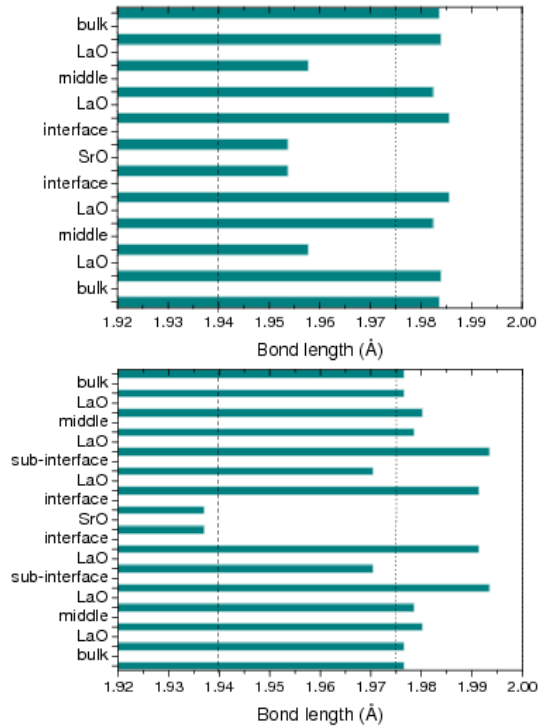} 
\caption{(Color online) V-O and O-V bond lengths along the $z$-direction for 
$m=5$ (top) and $m=6$ (bottom). The vertical dashed line indicates $a/2$.} 
\label{bond}
\end{figure}

The $m=5$ superlattice hosts different physics. Again,
lattice relaxation plays an important role. The obtained buckling of the
interface layers is summarized in Table~\ref{tab:1}.
The interface VO$_2$ layer shows a buckling with the same displacement
directions as for $m=6$ (the V ions move towards the SrO layer and the O ions
towards the LaO layer) but the amplitude is smaller. Notably, as shown in
Fig.~\ref{bond}, while for the $m = 6$ case the distance between the V ion in
the interface layer and the O ion in the SrO layer becomes smaller than $a$/2,
inducing the inversion of the orbital occupancy, in the $m = 5$ case the bond
lengths stay larger than $a$/2 in all the VO$_2$ layers. The smaller amplitude
of the buckling therefore does not induce an inversion of the orbital
occupation and the confinement of the charge carriers is clearly less effective than 
in the even case. Furthermore, the reduction of the bond lengths reaches
further into the LaVO$_3$ layer, where the bond length of the V ion with the
second LaO layer counted from the SrO layer is still strongly reduced. In
comparison, in the $m = 6$ case the same bond length is larger than $c$/2.

\begin{figure}[!b]
\includegraphics[width=0.45\textwidth]{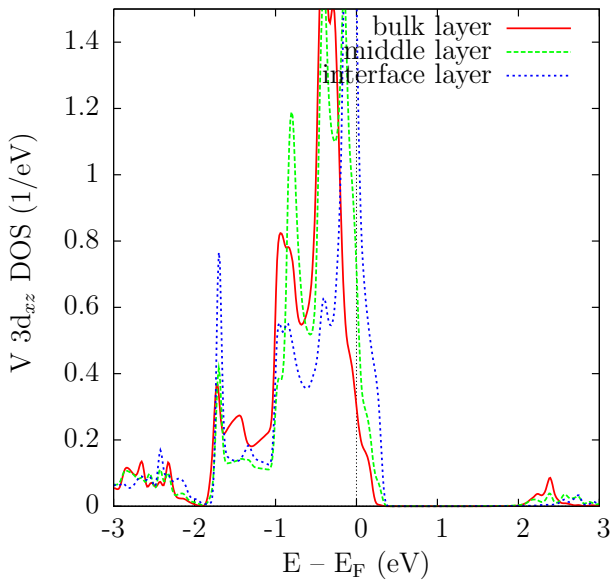}
\caption{(Color online) Spin-majority density of states projected on the V
$3d_{xz}$ orbital. The labeling of the layers is shown in Fig.~\ref{fig2}.}
\label{fig4}
\end{figure}

The change of the structural relaxation is visible in the resulting
layer-projected densities of states in 
Fig.~\ref{fig4}. They consist of the spin-majority density of states, as no
sizeable contribution from the spin-minority bands is obtained. All layers are
clearly off stoichiometry and all V ions are in a mixed valency state, yet
closer to the 3+ valency in the bulk layers than in the interface
layers. Moreover, all V $3d$ electrons occupy the $3d_{xz}$ and $3d_{yz}$
orbitals, while all $3d_{xy}$ orbitals are empty, in contrast to the $m=6$
superlattice.

\begin{figure}[!b]
\includegraphics[width=0.45\textwidth]{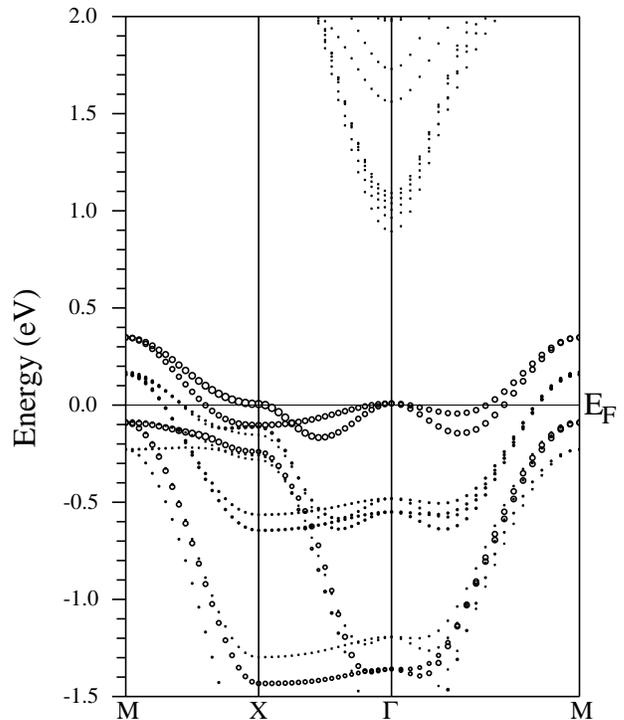}  
\caption{Band structure of the $m=5$ superlattice with ferromagnetically
polarized interfaces. The interface V $3d_{xz}$+$3d_{yz}$ orbitals are highlighted.} 
\label{fig5}
\end{figure}

The band structure of the $m=5$ superlattice is shown in
Fig.~\ref{fig5}. As there are no spin-minority bands close to the Fermi level, only
spin-majority bands are shown. There are two narrow bands formed by the interface V
$3d_{xz,yz}$ orbitals around the Fermi level. The four next lower lying bands 
are due to the middle and bulk layers with a larger band width.
The bands at the Fermi level are separated by a band gap of about 1 eV 
(measured at the $\Gamma$ point) from the parabolic bands of the bulk
states. In addition, the bands with the largest Fermi velocities mostly trace
back to the middle layer V $3d_{xz,yz}$ orbitals, with 
some admixtures of the bulk and interface layers. Therefore, transport is no
longer confined to the interface layers but there is a distinct
modulation associated to it. There are fast charge carriers mostly
confined to the next-to-interface layer, with a small admixture of
their two neighbors, and slow charge carriers mostly confined to the interface layer. 

Narrow $d$ bands with high DOS at the Fermi level often explain
ferromagnetism in the $3d$ transition metal oxides. Here the
high DOS follows from the slow charge carriers that are confined to the
interface layers. We therefore have performed calculations for $2\times2$
supercells in the $ab$-plane and find that the ground state shows a
ferromagnetic alignment of the spins within the interface layer, whereas they
are aligned antiferromagnetically in the other layers.

\section{Conclusion and outlook}\label{sec:concl}

Our calculations show that the doping resulting from the charge carriers
donated by the SrO layers in (LaVO$_3$)$_m$/SrVO$_3$ superlattices leads to 
rich physics built on fast and slow charge carriers together with lattice
relaxation. For both types of superlattices the lattice relaxation plays an
important role, demonstrated by the obtained buckling of the VO$_2$ interface
layers and alternation of short and long V-O bond lengths along the
$c$-axis. Although the buckling of the interface VO$_2$ layers is qualitatively
the same in the two studied cases, the critical difference in the amplitude
leads to two different band structures. In the odd $m = 5$ case, slow charge
carriers are mostly confined to the interface layers in which they order
ferromagnetically. The confinement of the fast charge carriers is less effective
and their distribution peaks in the middle layers. For even superlattices the
relaxation confines the fast charge carriers to largely decoupled interface
layers in which they form an ultrathin two-dimensional electron gas. No
magnetic order is found and the distribution of slow charge carriers peaks in the
sub-interface layers. Therefore, coming back to our original question, it seems unlikely
that the geometric confinement of the charge carriers alone is the origin of
the experimentally observed ferromagnetic state. Instead, the fact that the
even and odd superlattices show a different behavior points out the relevance
of the number of unit cells of the LaVO$_3$ layers.

In our calculations we did not take into account the orthorhombic structure of
LaVO$_3$,\cite{Bordet93} generated by the octahedral tilts which were shown to
be present in pure LaVO$_3$ thin films as well.\cite{Rotella12} The
related unit cell is larger than the tetragonal cell used in 
the calculations, in particular along the tetragonal $c$-axis, where the
out of phase rotation of the octahedra leads to a doubling of the unit cell, see
Fig.~\ref{fig2}. Therefore, in an even superlattice it is
possible to establish the bulk structure including the octahedral rotations in
the LaVO$_3$ layer. On the other hand, in the odd superlattice the thickness of
the LaVO$_3$ layer corresponds to a half-integer number of the orthorhombic unit
cell, so that the bulk structure (and therefore the bulk physics) of the
LaVO$_3$ layer cannot be established. The compensation by the structure of the
lack of the energetically favorable out of phase rotation may lead to the
small but critical differences of the lattice relaxations and therefore the
orbital occupations between the even and odd case. The combination of the
individual reactions of the structural, orbital, and electronic degrees of freedom
leads in the end to a ferromagnetic ground state in the odd superlattices. 

Our calculations confirm therefore the experimentally observed magnetization
oscillations,\cite{Lueders} although in the experiment the even superlattices
were found to show a higher magnetization than the odd ones. This difference
can be explained by the fact that the thickness of the LaVO$_3$ in the
experimental study is a nominal one, where the announced thickness does not
take into account any intermixing of La and Sr during the growth. It was shown
that the interface between the LaVO$_3$ and SrVO$_3$ layers is abrupt but
exhibits steps of typically one perovskite unit cell.\cite{Boullay11} The steps
lead to the presence of Sr atoms in the first LaO layer next to the SrO
layer. Taking into account the growth direction, it can be argued that the
thickness of the pure LaVO$_3$ layer will thus be one unit cell less than the
nominal one, resulting in an inversion of the nominally odd and nominally even
superlattices.

In conclusion, two scenarios for the even and odd superlattice have been
found. In both cases, the  $d_{xz}$ and $d_{yz}$ orbitals form bands crossing
the Fermi level. In the even superlattice, the orbital occupancy changes from
the interface to the bulk layers, while we find an homogeneous orbital occupation
in the odd superlattice. The room temperature magnetism is related to the incomplete
formation of the orthorhombic unit cell of LaVO$_3$ in odd superlattices. A
next step could be a comparison of theoretical and experimental macroscopic
transport properties of the confined fast charge carriers. 

\begin{acknowledgments} 
We thank S.\ Nazir for valuable discussions. Financial support by the German Research
foundation within TRR 80 (CS) and by the ANR through GeCoDo-ANR-2011-JS0800101 (UL and RF) is
acknowledged. UL and RF gratefully acknowledge the R\'egion Basse-Normandie
and the Minist\`ere de la Recherche for financial support. 
\end{acknowledgments}


\begin{thebibliography}{99}
\bibitem{Malozemoff05} A.\,P. Malozemoff, J.~Mannhart, and D.~Scalapino,
  Phys. Today \textbf{58}, 41 (2005).

\bibitem{Bed86}J.\,G. Bednorz and K.\,A. M\"uller, Z.\ Physik B \textbf{64},
  189 (1986); B.~Raveau, C.~Michel, M.~Hervieu, and D.~Groult, Crystal
  Chemistry of High-T$_c$ Superconducting Copper Oxides, Springer Series in
  Material Science \textbf{15}, Springer-Verlag Berlin, Heidelberg, New York
  (1991). 

\bibitem{Kawa97}H.~Kawazoe, H.~Yasakuwa, H.~Hyodo, M.~Kurota, H.~Yanagi, and H.~Hosono, Nature \textbf{389}, 939 (1997).

\bibitem{Li11} L.~Li, C.~Richter, S.~Paetel, T.~Kopp, J.~Mannhart, and
  R.~C.~Ashoori, Science {\bf 332}, 825 (2011).

\bibitem{Maignan09} I.\ Terasaki, Y.\ Sasago, and K.\ Uchinokura, \prb \textbf{56} 12 685 (R)  (1997); A.\ Maignan, V.\ Eyert, C.\ Martin, S.\ Kremer,
    R.\ Fr\'esard, and D.\ Pelloquin, Phys. Rev. B {\bf 80}, 115103 (2009).

\bibitem{Ohtomo04} A. Ohtomo and H.\ Y.\ Wang, Nature (London) {\bf 427}, 423
  (2004).  

\bibitem{thiel06} S.\ Thiel, G.\ Hammerl, A.\ Schmehl, C.\ W.\ Schneider, and
  J.\ Mannhart, Science {\bf 313}, 1942 (2006).

\bibitem{Shalom10} M.\ Ben Shalom, A.\ Ron, A.\ Palevski, and Y.\ Dagan, \prl
  {\bf 105}, 206401 (2010).

\bibitem{Kim11} M.\ Kim, C.\ Bell, Y.\ Kozuka, M.\ Kurita, Y.\ Hikita, and H.\
  Y.\ Hwang \prl {\bf 107}, 106801 (2011).

\bibitem{Jang11} H.\ W.\ Jang, D.\ A.\ Felker, C.\ W.\ Bark, Y.\ Wang, M.\ K.\
  Niranjan, et al, Science {\bf 331}, 886 (2011). 

\bibitem{Santander} A.\ F.\ Santander-Syro et al., Nature (London) {\bf  469},
  189 (2011). 

\bibitem{rey07} N.\ Reyren et al., Science {\bf 317}, 1196 (2007).

\bibitem{Lib11} L.~Li, C.~Richter, J.~Mannhart, and R.~C.~Ashoori, Nature
  Phys. {\bf 7}, 762 (2011). 

\bibitem{David11} A.\ David, R.\ Fr\'esard, Ph.\ Boullay, W.\ Prellier, U.\
  L\"uders, and P.-E.\ Janolin, \apl 98, 212103 (2011).
 
\bibitem{Lueders} U.\ L\"uders, W.\ C.\ Sheets, A.\ David, W.\ Prellier, and
  R.\ Fr\'esard, Phys.\ Rev.\ B {\bf 80}, 241102R (2009). 

\bibitem{Inaba} F.\ Inaba, T.\ Arima, T.\ Ishikawa, T.\ Katsufuji, and Y.\
  Tokura, Phys. Rev. B \textbf{52}, R2221 (1995).

\bibitem{Miyasaka00} S.\ Miyasaka, T.\ Okuda, and Y.\ Tokura, \prl
  \textbf{85}, 5388 (2000). 

\bibitem{Chan09} C.-K.\ Chan, P.\ Werner, and A.\ J.\ Millis, Phys.\ Rev.\ B
  \textbf{80}, 235114 (2009). 

\bibitem{Fresard02} R.\ Fr\'esard and M.\ Lamboley, J.\ Low
  Temp.\ Phys.\ \textbf{126}, 1091 (2002). 

\bibitem{Jackeli08} G.\ Jackeli and G.\ Khaliullin, Phys.\ Rev.\
  Lett. \textbf{101}, 216804 (2008). 
\bibitem{pickett10} V.\ Pardo and W.\ E.\ Pickett, Phys.\ Rev.\ B \textbf{81},
  245117 (2010). 

\bibitem{weng10} H.\ Weng and K.\ Terakura, Phys.\ Rev.\ B \textbf{82}, 115105
  (2010). 

\bibitem{Zubkov} G.\ Khaliullin, P.\ Horsch, and A.\ M.\ Ole\'{s}, 
\prl {\bf 86}, 3879 (2001); \prb {\bf 70}, 195103 (2004). 

\bibitem{fujika}
J.\ Fujioka, T.\ Yasue, S.\ Miyasaka, Y.\ Yamasaki, T.\ Arima, H.\ Sagayama,
T.\ Inami, K.\ Ishii, and Y.\ Tokura, Phys.\ Rev.\ B \textbf{82}, 144425 (2010). 

\bibitem{Anisimov} V.\ I.\ Anisimov, F.\ Aryasetiawan, and A.\ I.\ Lichtenstein,
J. Phys.: Condens. Matter \textbf{9}, 767 (1997).

\bibitem{schwin07} U.\ Schwingenschl\"ogl and C.\ Schuster,
Phys. Rev.  Lett. {\bf  99}, 237206 (2007).

\bibitem{nickelates} U.\ Schwingenschl\"ogl, C.\ Schuster, and R.\ Fr\'esard,
EPL \textbf{88}, 67008 (2009).

\bibitem{wien2k}
P.\ Blaha, K.\ Schwarz, G.\ Madsen, D.\ Kvasicka, and J.\ Luitz,
{\it Wien2k: An augmented plane wave + local orbitals program for calculating
  crystal properties}, Vienna University of Technology (2001).


\bibitem{inoue} I.\ H.\ Inoue, O.\ Goto, H.\ Makino, N.\ E.\ Hussey, and M.\
  Ishikawa, Phys. Rev. B \textbf{58}, 4372 (1998).

\bibitem{Bordet93} P.\ Bordet, C.\ Chaillout, M.\ Marezio, Q.\ Huang, A.\
  Santoro, S.\ W.\ Cheong, H.\ Takagi, C.\ S.\ Oglesby, and B.\ Batlogg, 
J. Sol. State Chem. \textbf{106}, 253 (1993).

\bibitem{Rotella12} H.~Rotella, U.\ L\"uders, P.-E.\ Janolin, V. H.\ Dao, D.\
  Chateigner, R.\ Feyerherm, E.\ Dudzik, and W.\ Prellier, Phys. Rev. B \textbf{85}, 184101 (2012).

\bibitem{Boullay11} P.\ Boullay, A.\ David, W.\ C.\ Sheets, U.\ L\"uders, W.\
  Prellier, H.\ Tan, J.\ Verbeeck, G.\ Van Tendeloo, C.\ Gatel, G.\ Vincze,
  and Z.\ Radi, Phys. Rev. B \textbf{83}, 125403 (2011).
\end{thebibliography}
\end{document}